\documentclass{ifacconf}

\usepackage{graphicx}      
\graphicspath{{figures/}}
\usepackage{natbib}        
\usepackage{amsmath, amssymb, amsfonts, mathtools}
\usepackage{booktabs}
\usepackage{subcaption}
\usepackage{color}
\usepackage{xurl}          
\begin{document}
\raggedbottom
\begin{frontmatter}

\title{Market Efficiency and Capacity Investment under Flexible Price-Making Loads:\\ A Computational Study} 

\author[MIT]{Aron Brenner} 
\author[MIT]{Deepjyoti Deka} 
\author[MIT]{Saurabh Amin}

\address[MIT]{Massachusetts Institute of Technology, 
   Cambridge, MA 02139 USA (e-mail: abrenner@mit.edu).}
   
\begin{abstract}                
Large flexible loads such as data centers can shift consumption across locations and hours to lower their bills. We model a price-making consumer leading the system operator in a Stackelberg game and compare its shifts with a cost-minimizing planner's over a simulated RTS-GMLC year. Strategic temporal and spatial shifting capture 75\% and 93\% of the planner's savings, forgoing the rest mostly by withholding flexibility to protect favorable prices. Scarce capacity and abundant renewables raise this share for temporal shifting; scarce transmission lowers it for spatial shifting. Temporal and spatial flexibility substitute for thermal and transmission capacity, respectively, and complement renewables.
\end{abstract}


\end{frontmatter}

\section{Introduction}
Rapid expansion of data centers is driving electricity load growth in the United States, with these facilities projected to account for up to 12\% of national consumption by 2028~\citep{shehabi20242024}. Despite strong pressure to bring these loads online, interconnection remains slow, as utilities and system operators struggle to expand transmission and generation capacity quickly enough to serve them~\citep{farmer2025doe}. Flexibility, such as data centers' ability to shift computation across time and location, has been proposed as a substitute for some of these capacity upgrades and, in turn, as a key path to faster interconnection~\citep{farmer2025doe, norris2025rethinking}.

Much research has investigated the potential for data center flexibility to reduce both electricity costs~\citep{qureshi2009, liu2011, zhang2020flexibility, knittel2025} and carbon emissions~\citep{liu2011, ZhengEtal2020_joule, LindbergEtal2022, radovanovic2022carbon, knittel2025}. These works evaluate flexibility's benefits either by letting a central planner choose the shift to minimize a system objective or by modeling the data center as a \textit{price-taking} consumer that minimizes its own cost or carbon without accounting for its effect on market prices. In both cases the flexible load's incentives are assumed to align with system efficiency, raising the question of how a large, \textit{price-making} consumer that shifts load \textit{strategically} to minimize its own bill affects system-level operations.

In this work, we build on \citet{brenner2026strategicspatialloadshifting}, which models strategic spatial load shifting as a bilevel program, and extend it to study two channels for flexibility: spatial and temporal load shifting. We evaluate the resulting strategic equilibria through simulation on the RTS-GMLC system~\citep{rts}, a synthetic 73-bus transmission network with a detailed generator fleet and high renewable penetration, built to represent a modern regional electricity system. We focus on the operational value of flexibility, the reduction in dispatch cost from shifting load on a system of fixed capacity, rather than the avoided or deferred capacity upgrades emphasized by \citet{farmer2025doe} and \citet{norris2025rethinking}. Our capacity experiments then show how this operational value changes under both centrally and privately optimal shifting as transmission, thermal, and renewable capacity vary. Our experiments address the following questions:
\begin{enumerate}
    \item How much of the system-cost reduction achievable under centralized coordination does privately optimal load shifting deliver?
    \item How does this efficiency differ between the spatial and temporal flexibility channels, and what failure mode is characteristic of each?
    \item How do changes in transmission, thermal, and renewable capacity affect the value of flexibility and the efficiency of strategic shifting?
\end{enumerate}
We proceed by presenting our bilevel load shifting model (Sec.~\ref{sec:model}) followed by experimental findings for the RTS-GMLC system (Sec.~\ref{sec:res}).

\section{Strategic Load Shifting Model}
\label{sec:model}

We consider a single strategic flexible consumer (e.g., a data center operator controlling several sites) that shifts its load in space and time to minimize its own electricity bill, anticipating how a wholesale electricity market clears. We model this interaction as a Stackelberg game, in which the flexible consumer leads with its load shift and the system operator follows by clearing the market through economic dispatch, and compute its equilibrium by solving a bilevel program. Throughout, $\mathcal B$, $\mathcal L$, $\mathcal G$, and $\mathcal T$ denote the buses, lines, generators, and time periods. The consumer has nominal load $\bar q_{b,t}$ at bus $b$ in period $t$ and chooses a shift $\delta_{b,t}$, so its load is $q_{b,t}=\bar q_{b,t}+\delta_{b,t}$.

\subsection{Lower Level: DC-OPF Market Clearing}
\label{subsec:ll}

We first present the system operator's (i.e., follower's) economic dispatch problem used to clear the electricity market. Here, we assume a linear DC optimal power flow (DC-OPF) model solved independently for each period with time-varying renewable availability ($\bar P_{g,t}$ for renewable resources $g$) and inflexible demand $D_{b,t}$. We stack the per-bus, per-line, and per-generator quantities into vectors, with $c$ the generator marginal costs and $p_t$ the dispatch, and collect the network structure in the bus--line incidence matrix $A\in\{0,\pm1\}^{|\mathcal B|\times|\mathcal L|}$ ($+1$ at a line's sending bus, $-1$ at its receiving bus), the bus--generator incidence matrix $C\in\{0,1\}^{|\mathcal B|\times|\mathcal G|}$, and the diagonal susceptance matrix $B_f$. For a given shift $\delta_t$, total demand is $d_t = D_t + \bar q_t + \delta_t$, and the system operator solves
\begin{subequations}\label{eq:ll}
\begin{align}
    \min_{p_t,\theta_t}\ \ & c^{\top} p_t \label{eq:ll-obj}\\
    \mathrm{s.t.}\ \ & C p_t - A f_t = d_t & (\lambda_t) \label{eq:ll-bal}\\
    & {-s} \le f_t \le s & (\mu^{\pm}_t) \label{eq:ll-flow}\\
    & 0 \le p_t \le \bar P_t & (\eta_t) \label{eq:ll-gen}\\
    & \theta_{b_0,t}=0, \label{eq:ll-slack}
\end{align}
\end{subequations}
where $f_t = B_f A^{\top}\theta_t$ is the vector of DC line flows. Constraint \eqref{eq:ll-bal} is the nodal balance, in which generation $C p_t$ plus net line inflow $-A f_t$ meets demand $d_t$. Constraint \eqref{eq:ll-flow} enforces the line thermal limits $s$, \eqref{eq:ll-gen} the generator capacities, and \eqref{eq:ll-slack} fixes the angle $\theta_{b_0,t}$ of slack bus $b_0$.
The system operator's problem has no intertemporal constraints. We do not model unit commitment, whose binary decisions would preclude the duality-based single-level reformulation of Sec.~\ref{subsec:solution}. Omitting ramp limits is a mild simplification at hourly resolution, since every thermal unit in RTS-GMLC can traverse its full output range within an hour (coal units in at most 53 minutes, combined-cycle units in 45).
\subsection{Upper Level: Strategic Flexible Consumer}

We now present the flexible consumer's (i.e., leader's) load shifting problem. Here, the flexible consumer chooses the shift $\delta$ to minimize its own procurement cost (its bill),
\begin{equation}\label{eq:ul-obj}
\min_{\delta \in \Delta}\ \ \sum_{t\in\mathcal{T}} w_t \sum_{b\in\mathcal{B}}
   \lambda_{b,t} q_{b,t},
\end{equation}
where $\Delta$ is the feasible shift set with an energy-con\-ser\-va\-tion constraint and bounds set by the shiftable fraction $\alpha$,
\begin{align}\label{eq:ul-con}
    \Delta \coloneqq \Big\{\delta \in \mathbb{R}^{|\mathcal B| \times |\mathcal{T}|} \, : \, & \sum_{b \in \mathcal B} \sum_{t \in \mathcal T} w_t\delta_{b,t}=0, \notag \\
    & {-\alpha}\bar q_{b,t}\le \delta_{b,t}\le \alpha\bar q_{b,t} \; \forall b,t \Big\}.
\end{align}
Here, each weight $w_t$ is the number of hours in temporal block $t$, so the weighted sums in \eqref{eq:ul-obj}--\eqref{eq:ul-con} approximate daily totals. The prices $\lambda_t$, i.e., the locational marginal prices (LMPs), are the duals of the balance constraint \eqref{eq:ll-bal} at the lower-level optimum under the shifted demand $d_t(\delta)$: the consumer anticipates how its shift re-prices the market, so \eqref{eq:ul-obj}--\eqref{eq:ul-con} with \eqref{eq:ll} embedded for every $t$ is a bilevel program.

\begin{figure}[tb]
\centering
\begin{subfigure}[t]{0.49\columnwidth}
\centering
\includegraphics[width=\linewidth]{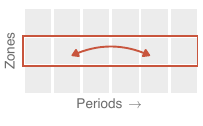}
\caption{Temporal: load moves within a zone, across periods.}
\label{fig:concept-temporal}
\end{subfigure}\hfill
\begin{subfigure}[t]{0.49\columnwidth}
\centering
\includegraphics[width=\linewidth]{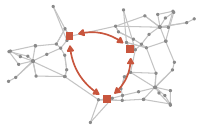}
\caption{Spatial: load moves within a period, across buses.}
\label{fig:concept-spatial}
\end{subfigure}
\caption{The two flexibility channels. Temporal shifting conserves each zone's daily energy, and spatial shifting conserves total load in each period. Squares mark the flexible-load buses (118, 218, and 318).}
\label{fig:concept}
\end{figure}

\textbf{Spatial vs.\ temporal shifting.}
We study two physically distinct flexibility channels, distinguished by the period set $\mathcal{T}$ and by the energy-conservation constraint that replaces the aggregate balance in \eqref{eq:ul-con}:
\begin{itemize}
\item \textit{Temporal} ($|\mathcal{T}|>1$): $\sum_t w_t\delta_{b,t}=0$ for each bus $b$ (a zone in our 3-zone experiments). Load is rescheduled within a bus, deferring or advancing consumption across the periods of the day while each site's daily energy is conserved.
\item \textit{Spatial} ($|\mathcal{T}|=1$): $\sum_b \delta_{b,t}=0$. Load is rerouted across buses, migrating consumption between sites within a period while the system-wide total is conserved.
\end{itemize}
For a data center, the temporal channel corresponds to rescheduling deferrable computation and the spatial channel to migrating workload between sites. Fig.~\ref{fig:concept} illustrates the two channels. In the temporal channel (Fig.~\ref{fig:concept-temporal}), each zone's load moves along its own time axis across the day's temporal blocks. In the spatial channel (Fig.~\ref{fig:concept-spatial}), load moves between the flexible-load buses of the network within a single period, so its value and its strategic opportunities derive entirely from network congestion.

\textbf{Centralized benchmark.}
To measure the efficiency of strategic shifting, we compare it against a centralized benchmark in which a planner chooses the shift to minimize total dispatch cost over the same feasible set. We write the dispatch cost as
\begin{equation}\label{eq:soc}
V(\delta) = \sum_{t\in\mathcal{T}} w_t \sum_{g\in\mathcal{G}} c_g p_{g,t}(\delta),
\end{equation}
where $p_{g,t}(\delta)$ is the dispatch when the market \eqref{eq:ll} clears under $\delta$. With $\Delta$ as in \eqref{eq:ul-con}, the \textit{centrally optimal} shift is
\begin{equation}\label{eq:central}
    \delta^c \in \arg\min_{\delta\in\Delta} V(\delta),
\end{equation}
which is a linear program with the same shift bounds, periods, and market model as the strategic problem. We denote by $S^c \coloneqq V(0)-V(\delta^c)$ the \textit{value of flexibility}, the planner's savings in dispatch cost, and by $S^\star \coloneqq V(0)-V(\delta^\star)$ the \textit{strategic savings} under the strategic shift $\delta^\star$ solving \eqref{eq:ul-obj}. We measure the efficiency loss from strategic behavior by the \textit{efficiency gap}, the value lost relative to the planner,
\begin{equation}\label{eq:gap}
    G \coloneqq S^c - S^\star = V(\delta^\star) - V(\delta^c)\ge 0,
\end{equation}
and report the \textit{efficiency} $S^\star/S^c$. Strategic shifting harms the system outright ($S^\star<0$) exactly when the gap exceeds the value of flexibility.

\subsection{Single-Level Reformulation and Solution}
\label{subsec:solution}

Because the lower level \eqref{eq:ll} is a linear program, it can be replaced exactly by its primal feasibility, dual feasibility, and strong-duality conditions~\citep{gabriel2013complementarity}. Dual feasibility is given by the constraints
\begin{subequations}\label{eq:dual}
\begin{align}
& C^{\top}\lambda_t-\eta_t\le c,
  \quad \eta_t,\mu^{+}_t,\mu^{-}_t\ge 0,
  \label{eq:dual-gen}\\
& \big(A\xi_t\big)_b = 0
  \quad \forall b\neq b_0,
  \label{eq:dual-ang}
\end{align}
\end{subequations}
where $\xi_t\coloneqq B_f\big(A^{\top}\lambda_t+\mu^{+}_t-\mu^{-}_t\big)$ is the vector of dual line flows, so that \eqref{eq:dual-ang} is the stationarity condition for $\theta_t$. Then, the strong-duality condition is
\begin{equation}\label{eq:sd}
    c^{\top} p_t = \lambda_t^{\top} d_t - \eta_t^{\top}\bar P_t - s^{\top}(\mu^{+}_t+\mu^{-}_t).
\end{equation}
Stacking \eqref{eq:ll-bal}--\eqref{eq:ll-slack}, \eqref{eq:dual}, and \eqref{eq:sd} for all $t$ as constraints of \eqref{eq:ul-obj}--\eqref{eq:ul-con} yields an equivalent single-level program in which the prices are decision variables constrained to be optimal duals of the dispatch. The problem is nonconvex because products of prices and the shift appear in the objective \eqref{eq:ul-obj} and through $\lambda_t^{\top}d_t$ in \eqref{eq:sd}. We solve it with Gurobi's spatial branch-and-bound \citep{gurobi}.

\section{RTS-GMLC Simulation}
\label{sec:res}
\subsection{Experimental Setup}

The bilevel program of Sec.~\ref{sec:model} has no closed-form solution, and its outcomes vary with load, renewable availability, and installed capacity. We therefore evaluate it computationally over one year (366 days) of hourly load and renewable availability on the 73-bus RTS-GMLC system. We consider two representations: a 3-zone aggregation that keeps individual generators and sums inter-zonal transmission capacity, which we use to model temporal shifting, and the full 73-bus network, used to model spatial shifting. We assume the flexible consumer has a constant load of 250~MW in each zone (750~MW in total, or about 15\% of annual system energy), of which a fraction $\alpha=0.25$ can be shifted at each bus in each period. On the 73-bus network, we site the flexible load at the largest-load bus in each zone.

For tractability, we divide each day into six blocks of consecutive hours that follow the daily net-load curve: overnight (hours 0--5), the morning down-ramp (hours 5--9), the midday solar trough (hours 9--15), the afternoon up-ramp (hours 15--19), the evening (hours 19--22), and late evening (hours 22--24). The morning and midday blocks contain the daily net-load minimum on $90\%$ of days, and the afternoon and evening blocks contain the peak on $89\%$. Each block is represented by the hour whose net load is closest to the block's mean, except the block containing the daily peak, which is represented by the peak hour. Representative hours are chosen once, for the base system (RTS-GMLC as published), and held fixed across the capacity scenarios of Sec.~\ref{subsec:res-invest}. The temporal channel solves the six blocks jointly, while the spatial channel's per-period balance decouples them into six independent rerouting problems.\footnote{Each strategic problem has a 300-second limit, and we keep a day only if it solves to optimality with a relative gap of at most $10^{-3}$, which holds for 363 of 366 days in the temporal channel and 365 in the spatial channel in the base system.}

\subsection{Value and Efficiency over the Year}
\label{subsec:res-when}

\begin{figure}[tb]
\centering
\includegraphics[width=\columnwidth]{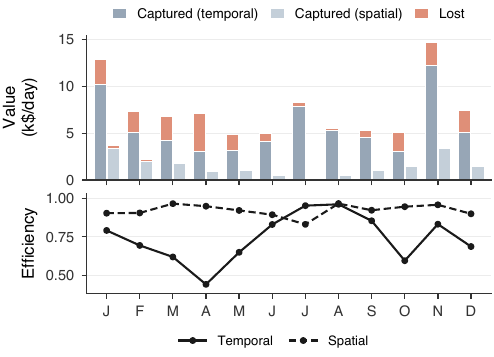}
\caption{Top: mean daily value of flexibility $S^c$ by month, split into the part captured by strategic shifting ($S^\star$) and the part lost ($G$). Bottom: monthly efficiency.}
\label{fig:seasonal}
\end{figure}

Recall that the \textit{value of flexibility} $S^c$ is the dispatch-cost savings under the planner's centrally optimal shift $\delta^c$. It is a theoretical benchmark, the most that any feasible shift can save, and the \textit{efficiency} $S^\star/S^c$ is the share of it that strategic shifting captures. Over the full year, with about $4\%$ of system energy shiftable ($\alpha$ times the consumer's $15\%$), we find the maximum achievable savings to be \$2.7 million in the temporal channel and \$0.58 million in the spatial channel, or $0.45\%$ and $0.09\%$ of the roughly \$0.6 billion annual dispatch cost without shifting. Strategic shifting captures $75\%$ and $93\%$ of these savings, respectively.

\textbf{Seasonal variation.} Fig.~\ref{fig:seasonal} shows that both quantities vary over the year, but differently for temporal and spatial shifting. The value of temporal shifting persists all year but its efficiency is seasonal. Conversely, the value of spatial shifting is seasonal but its efficiency stays high. The efficiency of temporal shifting (Fig.~\ref{fig:seasonal}, bottom) is highest in July and August ($0.95$ and $0.96$) and lowest in the spring and fall shoulder seasons ($0.44$ in April and $0.59$ in October). Sec.~\ref{subsec:res-mech} proposes an explanation based on the merit-order curve.

The value of spatial shifting is highest in winter and lowest in summer (Fig.~\ref{fig:seasonal}, top), yet its monthly efficiency stays between $0.83$ and $0.97$. Seasonality thus affects how much spatial flexibility is worth rather than how well private incentives capture system-wide value. Spatial flexibility needs congestion to be valuable, since rerouting load pays only when a transmission limit opens a price spread between buses. In the RTS-GMLC system, these spreads appear to track renewable output. When renewables meet a large share of demand, their concentrated output creates local surpluses that line limits keep from reaching load elsewhere in the system, so spreads are largest in the low-net-load winter and shoulder months and smallest under summer's high, thermally served loads.

\subsection{Mechanisms of Inefficiency}
\label{subsec:res-mech}

\begin{figure}[tb]
\centering
\includegraphics[width=\columnwidth]{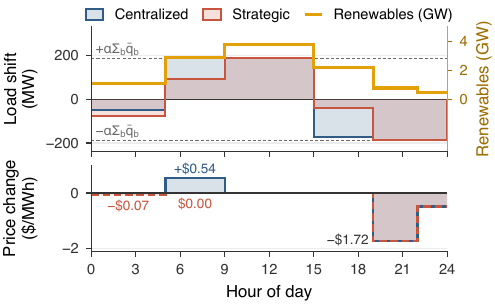}
\caption{Withholding flexibility on day 144 (temporal channel). Top: system-wide load shift and renewable output. Bottom: price change relative to no shifting.}
\label{fig:caseday}
\end{figure}

\begin{figure}[tb]
\centering
\includegraphics[width=\columnwidth]{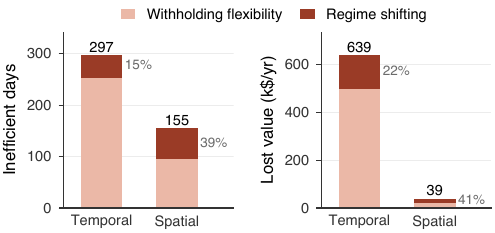}
\caption{Efficiency gap by failure mode (withholding flexibility or regime shifting), in inefficient days (left) and lost value (right). Percentages give the regime-shift share.}
\label{fig:regimebreakdown}
\end{figure}

Strategic shifting thus forgoes $25\%$ of the temporal value of flexibility but only $7\%$ of the spatial value. To explain this difference, and the seasonal pattern in Fig.~\ref{fig:seasonal}, we compare the strategic shift with the planner's on each day.

\textbf{Two failure modes.} We propose a decomposition of inefficient outcomes into two modes, distinguished by the \textit{active set}: the transmission lines and generators whose limits bind in the market clearing during any of the day's periods. We call a day inefficient if strategic shifting forgoes more than $0.1\%$ of the planner's savings, and we classify it by whether the strategic shift ends on the planner's active set, so the two modes are exhaustive and mutually exclusive. In both modes the consumer moves load to lower the prices it pays.

Under the first mode, \textit{withholding flexibility}, the consumer typically shifts in the planner's direction but halts at the edge of a merit-order step, where a cheap marginal unit reaches capacity. At the edge, the price is not unique: it can be the cheap unit's cost or the next unit's, and we assume the consumer pays the lower one on its entire load.\footnote{At such points the DC-OPF \eqref{eq:ll} is degenerate and the LMPs are not unique. As in \citet{brenner2026strategicspatialloadshifting}, we take the bill-minimizing optimal duals. This is without loss of generality, as they are the limit of arbitrarily small perturbations that restore uniqueness.} The planner, indifferent to prices, pushes past the edge, where the price is the higher one. The consumer thus reaches the planner's active set but deploys less flexibility. Under the second mode, \textit{regime shifting}, the strategic shift instead ends on a different active set than the planner's, which re-prices the system. The consumer makes a different set of lines or generators bind because the resulting prices lower its bill more than any shift that keeps the planner's active set.

\textbf{Why the consumer withholds.} Withholding flexibility, the source of most losses in both channels, is a form of monopsony behavior driven by the merit order. Concentrating load where power is cheap, such as a renewable-rich period or a bus near cheap generation, eventually exhausts that supply and lifts the price up a merit-order step. Because the consumer would pay the higher price on all the load it has already placed, it halts before the step. When the consumer's market power over constraints is more limited, which we test by setting $\alpha=0.1$, the planner's savings fall by more than half in both channels, but efficiency rises from $0.75$ to $0.85$ in the temporal channel and from $0.93$ to $0.94$ in the spatial channel.

\textit{Example.} Fig.~\ref{fig:caseday} illustrates the consumer withholding flexibility on one day of temporal shifting. The planner fills the morning net-load valley to its bound ($+187.5$~MW), replacing costly peak generation with cheap off-peak energy. The strategic consumer fills it only about halfway ($+92$~MW). Filling it completely would exhaust the cheap marginal unit and raise the valley price from \$24.50 to \$25.04/MWh, a price the consumer would pay on all of its valley load, not just the last increment. Withholding keeps the price at \$24.50 but leaves valley capacity unused. The strategic shift captures \$4{,}870 of the planner's \$6{,}050 savings and loses \$1{,}180, but cuts the consumer's bill by \$880 relative to the planner's shift.

\textbf{Losses by channel and mode.} Fig.~\ref{fig:regimebreakdown} splits the efficiency gap by mode. Beyond the larger temporal gap already visible in Fig.~\ref{fig:seasonal}, the channels differ in how they lose value: temporal losses mostly result from withholding flexibility, with regime shifts accounting for $15\%$ of inefficient days and $22\%$ of lost value. On the congested 73-bus network, spatial regime shifts are more common, accounting for $39\%$ of inefficient days and $41\%$ of lost value.\footnote{Because the two channels use different networks, we repeated the spatial experiment on the 3-zone system with the temporal channel's blocks, siting, and solver settings. The ranking holds (efficiency $0.96$ spatial versus $0.75$ temporal), although regime shifts then account for only $3\%$ of the spatial gap because aggregation removes most congestion.} Outright harm ($S^\star<0$) is rare, occurring on 22 days under temporal shifting and 10 days under spatial shifting. In the temporal channel, this usually involves a strategic shift that points against the planner's, since a shift that only scales down the planner's cannot raise dispatch cost.

\textbf{Why spatial shifting is more efficient.} The share of value lost in either channel is the product of \emph{exposure}, the share of the value of flexibility at stake in decisions where strategic shifting loses value (inefficient decisions, by the same $0.1\%$ test), and \emph{severity}, the share of that value it loses there. A decision spans a day in the temporal channel, which balances load over each day, and a period in the spatial channel, which balances it within each period. Severity is in fact higher in the spatial channel: inefficient periods lose $53\%$ of the value at stake (73-bus), versus $29\%$ for inefficient temporal days (3-zone). The difference lies in exposure. Only $14\%$ of spatial value lies in inefficient periods, compared with $86\%$ of temporal value in inefficient days ($0.14 \times 0.53 \approx 0.07$ versus $0.86 \times 0.29 \approx 0.25$ of value lost).

Exposure differs because strategic behavior requires a \emph{price breakpoint} within reach of the consumer's shift, that is, a shift level at which some line or generator starts or stops binding and prices jump. A temporal shift changes each period's total demand and so moves along the merit order directly. A spatial shift is net-zero within each period, so it moves prices only through congestion and rarely reaches a breakpoint. Across days, larger price spreads raise the stakes (i.e., the value of flexibility $S^c$) but not the share of it that is lost. On the same 3-zone system, intertemporal spreads average \$7.28/MWh and spatial spreads \$0.30/MWh (zero on $62\%$ of days), yet within each channel the share lost does not rise with the spread.

\begin{figure}[tb]
\centering
\includegraphics[width=\columnwidth]{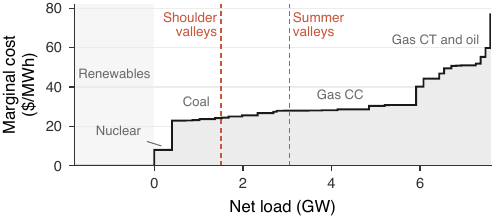}
\caption{Merit-order supply curve of the RTS-GMLC fleet against net load (load minus renewables). Negative net load is a renewable surplus. Dashed lines mark the median daily minimum net load (valley), including the flexible load, in the shoulder seasons and in summer. CC: combined cycle. CT: combustion turbine.}
\label{fig:meritorder}
\end{figure}

\textbf{Seasonal efficiency.} Withholding also accounts for when temporal shifting is efficient (Fig.~\ref{fig:seasonal}, bottom). The supply curve (Fig.~\ref{fig:meritorder}) has one large step, from renewables and nuclear (at most \$8/MWh) up to coal at about \$23/MWh, above which coal and gas CC form a gently rising stack up to about 6~GW. Because the gain from withholding scales with the price jump it avoids, efficiency depends on where the net-load valley, the period of lowest net load into which temporal shifting moves load, falls on this curve. In the high-load summer, even the valley is served well up the smooth thermal stack, where adjacent units differ by about \$1/MWh or less, so withholding gains the consumer relatively little and the strategic shift nearly matches the planner's. In the shoulder seasons, lower load and ample renewables push the valley down to this step, a jump of about \$15/MWh, on about a fifth of days. Withholding pays far more on these days, which carry two-thirds of the shoulder loss.

\subsection{Capacity Investment}
\label{subsec:res-invest}
\begin{figure}[tb]
\centering
\includegraphics[width=\columnwidth]{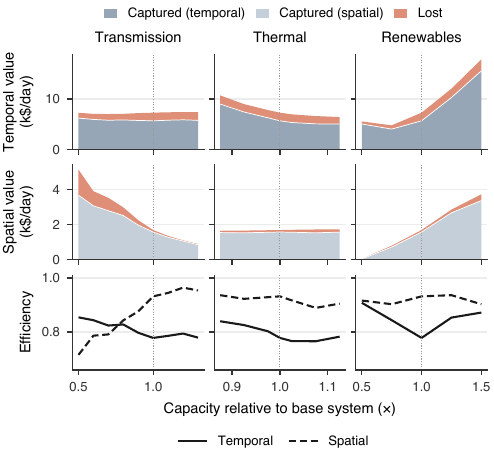}
\caption{Value of flexibility $S^c$ (rows 1--2), split into captured ($S^\star$) and lost ($G$) parts, and efficiency (row 3), versus line ratings, thermal capacity, and renewable availability relative to the base system. Note: averages are over days solved at every capacity level.}
\label{fig:effcap}
\end{figure}

Flexibility has been proposed as a substitute for transmission and generation upgrades \citep{farmer2025doe, norris2025rethinking}. To test whether it substitutes for or complements each technology operationally, and to understand how capacity changes the efficiency of strategic shifting, we vary the system-wide capacity of one technology at a time relative to the base system: all line ratings over $[0.5,1.3]$, all thermal capacity over $[0.875,1.125]$, and all renewable availability over $[0.5,1.5]$. Fig.~\ref{fig:effcap} reports temporal results on the 3-zone system and spatial results on the 73-bus network.

\textbf{Each channel substitutes for one form of capacity and complements renewables.} The value of flexibility (rows~1--2) mirrors the resource each channel exploits. Temporal value is essentially flat in transmission, falls by about $40\%$ from the scarcest to the most abundant thermal fleet, and more than triples as renewables are added. Rescheduling arbitrages intertemporal price spreads, which dispatchable thermal capacity flattens and renewables deepen. Spatial value, by contrast, falls by more than $80\%$ as line ratings rise from half to $1.3$ times their base values but is largely insensitive to thermal capacity. Temporal flexibility thus substitutes for thermal capacity and spatial flexibility for transmission.

Renewables raise the value of both channels, deepening net-load valleys for temporal shifting and creating congestion-driven price spreads for spatial shifting, so both channels complement renewable generation.

\textbf{Scarcity and abundant renewables raise temporal efficiency.} Temporal efficiency (row~3) rises as any form of capacity becomes scarcer and, along the renewables axis, also as renewables become abundant ($0.91$ when renewables are scarce, $0.78$ in the base system, and $0.87$ when abundant). Adding transmission or thermal capacity beyond the base system changes it little. A likely explanation is that strategic behavior requires slack in the dispatch: under scarcity, constraints bind regardless of the shift, so the consumer's private optimum tracks the planner's. Only with slack can the consumer choose whether a constraint binds and stop at a profitable price step. With abundant renewables, many valleys sit in surpluses deep enough to absorb the entire shift at zero price, leaving no step within reach, and these days carry much of the added value. Lost value can still grow as efficiency rises: with abundant renewables, the temporal loss rises from about \$1{,}600 to \$2{,}300 per day because the value at stake grows faster.

\textbf{Scarce transmission makes spatial shifting less efficient.} For spatial shifting, transmission has the opposite effect: halving line ratings triples its value but lowers its efficiency from $0.93$ to $0.72$, below that of temporal shifting. Tighter lines likely put more congestion breakpoints within the consumer's reach, raising its exposure (Sec.~\ref{subsec:res-mech}). Along the thermal and renewables axes, spatial efficiency stays between $0.89$ and $0.94$. Overall, strategic temporal shifting is most efficient in capacity-scarce or renewable-rich systems, whereas strategic spatial shifting is least efficient where transmission is scarce.

\section{Conclusion}
We studied how a large, price-making flexible load reshapes market outcomes when it shifts consumption to minimize its own bill. Over a simulated year of the RTS-GMLC system, strategic shifting realizes most of the planner's savings, but the remaining loss follows a consistent pattern: the consumer exploits the dispatch's active constraints, mostly by withholding flexibility at merit-order steps and sometimes by shifting which constraints bind, especially in the spatial channel. In the base system, these losses are larger for temporal than spatial shifting, because temporal shifts reach price breakpoints far more often. Temporal and spatial flexibility substitute for thermal and transmission capacity, respectively, and both complement renewables. Temporal efficiency rises when capacity is scarce or renewables are abundant, whereas spatial efficiency falls as transmission becomes scarce. Future work will develop a theoretical analysis of these phenomena and extend it to multiple strategic consumers.

\section*{DECLARATION OF GENERATIVE AI AND AI-ASSISTED TECHNOLOGIES IN THE WRITING PROCESS}
During the preparation of this work the authors used Claude Code to assist with writing code, generating figures, and drafting and editing text. After using this tool, the authors reviewed and edited the content as needed and take full responsibility for the content of the publication.

\bibliography{ifacconf}

\end{document}